\documentclass[epj,twocolumn]{webofc}
\usepackage[varg]{txfonts}   
\usepackage{hyperref}
\hypersetup{
 bookmarksnumbered=true,
 pdftitle = {What could we learn from a sharply falling positron fraction?}, 
 pdfauthor = {Timur Delahaye, Kumiko Kotera, Joseph Silk}, 
 pdfsubject = {positron fraction}, 
 pdfkeywords = {positron fraction, AMS-02, dark matter, cosmic rays},
 colorlinks=true,
 linkcolor=blue,
 citecolor=blue,
 urlcolor=blue,
}
\woctitle{SUGAR 2015}
\begin{document}
\title{What can we learn from a sharply falling positron fraction?}
%
%

\author{Timur Delahaye\inst{1}\fnsep\thanks{\email{timur.delahaye@polytechnique.org}} \and
        Kumiko Kotera\inst{2}\fnsep \and
        Joseph Silk\inst{2,3,4}\fnsep
}

\institute{Oskar Klein Centre for Cosmoparticle Physics, Department of Physics, Stockholm University, SE-10691 Stockholm, Sweden 
\and
           Institut d’Astrophysique de Paris UMR7095—CNRS, Université Pierre and Marie Curie, 98 bis Boulevard Arago, F-75014 Paris, France 
\and
          Department of Physics and Astronomy, The Johns Hopkins University, 3400 North Charles Street, Baltimore, MD 21218, USA
\and Beecroft Institute of Particle Astrophysics and Cosmology, Department of Physics, University of Oxford, DenysWilkinson Building, 1 Keble Road, Oxford OX1 3RH, UK
          }

\abstract{%
Recent results from the AMS-02 data have confirmed that the cosmic ray positron fraction increases with energy between 10 and 200GeV. This quantity should not exceed 50\%, and it is hence expected that it will either converge towards 50\% or fall. We study the possibility that future data may show the positron fraction dropping down abruptly to the level expected with only secondary production, and forecast the implications of such a feature in term of possible injection mechanisms that include both Dark Matter and pulsars.
}
\maketitle
\section*{Introduction}
\label{intro}
Since the publication of their results by the PAMELA collaboration~\citep{Adriani2009,Adriani2013}, the positron fraction \textit{i.e.} the flux of cosmic-ray positrons divided by the flux of electrons and positrons, has attracted a lot of interest. Indeed, PAMELA observed a raise of this quantity together with the cosmic-ray energy between 10 and 200~GeV which has been confirmed by AMS-02~\citep{Aguilar2013}. 

The AMS-02 experiment should have the ability to measure the positron fraction at even higher energies. What ever is the correct explanation for this rise, the positron fraction must either saturate or decline. In the latter case, how abrupt a decline might we expect? The AMS-02 collaboration is prone to explain that a sharply falling positron fraction would be a smoking gun for Dark Matter (see for instance the AMS-02 press conference of September 2014\footnote{\url{http://goo.gl/sf71o6}} or Manuela Vecchi's presentation at SUGAR 2015). In this work we studied whether this affirmation was motivated by any solid scientific argument.

Because stars do not contain anti-matter, positrons, like anti-protons or anti-deuterons, that we find in cosmic rays, are expected to be produced as {\it secondary} particles by cosmic ray nuclei while they propagate and interact in the interstellar medium (ISM). It is now clear that the increase observed in the positron fraction cannot be explained by the simplest models of secondary production. Various alternatives have been proposed, such as a modification of the propagation model~\citep{Katz2009,Blum2013}, or primary positron production scenarios, with pulsars~(\textit{e.g.}, \citealp{Grasso09,Hooper09,Delahaye2010,Blasi11,Linden13}) or Dark Matter annihilation~(\textit{e.g.}, \citealp{Delahaye2008,Arkani-Hamed09,Cholis09,Cirelli09}) as sources. As of today, it is not possible to conclude which explanation is the correct one because they all suffer from theoretical uncertainties and because the current data cannot lift degeneracies. Finding a way to discriminate among these explanations has been looked for by variuos authors (see for instance \citealp{Ioka2010,Kawanaka2010,Pato2010,Mauro2014}); here we want to test more specifically the possibility of a sharp drop of the positron fraction. An original aspect of our work is to also convolve our results with the cosmic-ray production parameter space for pulsars allowed by theory. We investigate the following question: what constraints could we put on Dark Matter annihilation and primary pulsar scenarios if the next AMS-02 data release were to show a sharply dropping positron fraction? 

According the AMS-02 collaboration, a sharp drop can only be explained if the positron excess originates from the annihilation of Dark Matter particles with a mass of several hundred GeV. However, we show here that such a feature would be highly constraining in terms of Dark Matter scenarios. 
In fact pulsar models could lead to a sharp fall of the positron fraction at the cost of some parameter tuning.

In this proceedings we summarize the results of our earlier work~\citep{Delahaye2014}; readers interested in the details of the method are advised to consult that earlier reference.

\section*{Positron flux morphology}
\label{sec-1}

\begin{figure*}
\centering
\includegraphics[width=\columnwidth,clip]{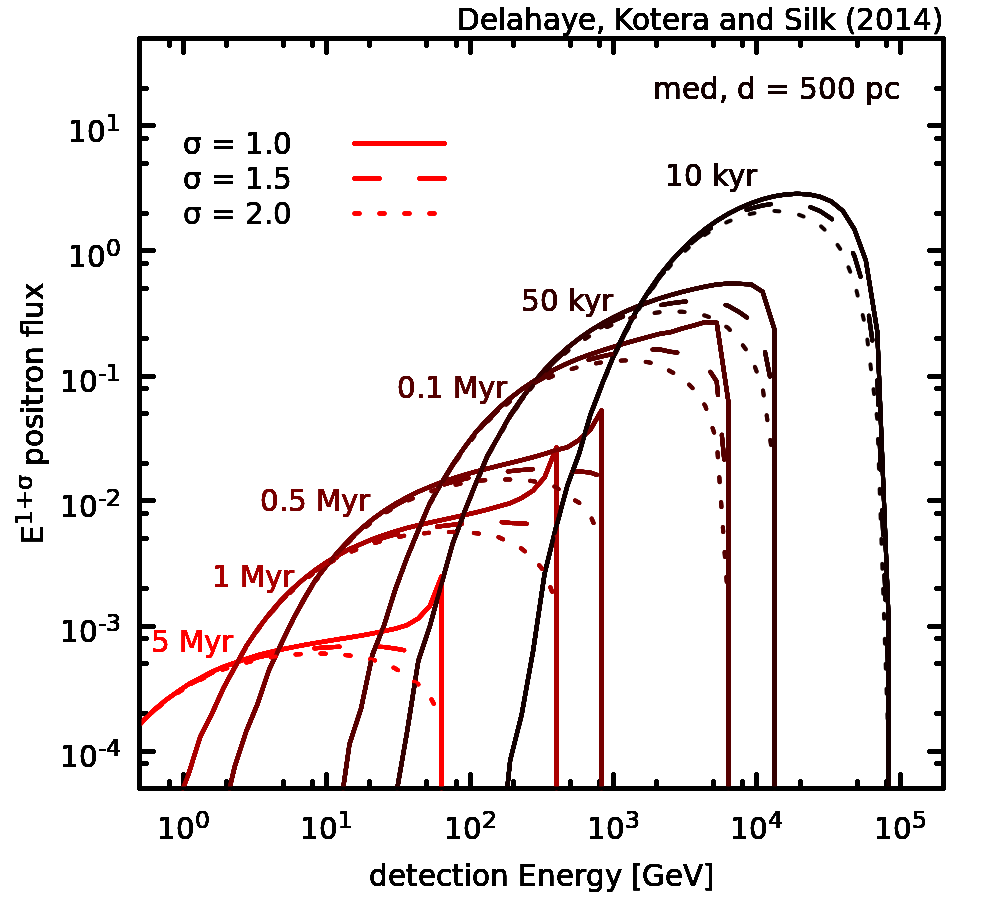}
\includegraphics[width=\columnwidth,clip]{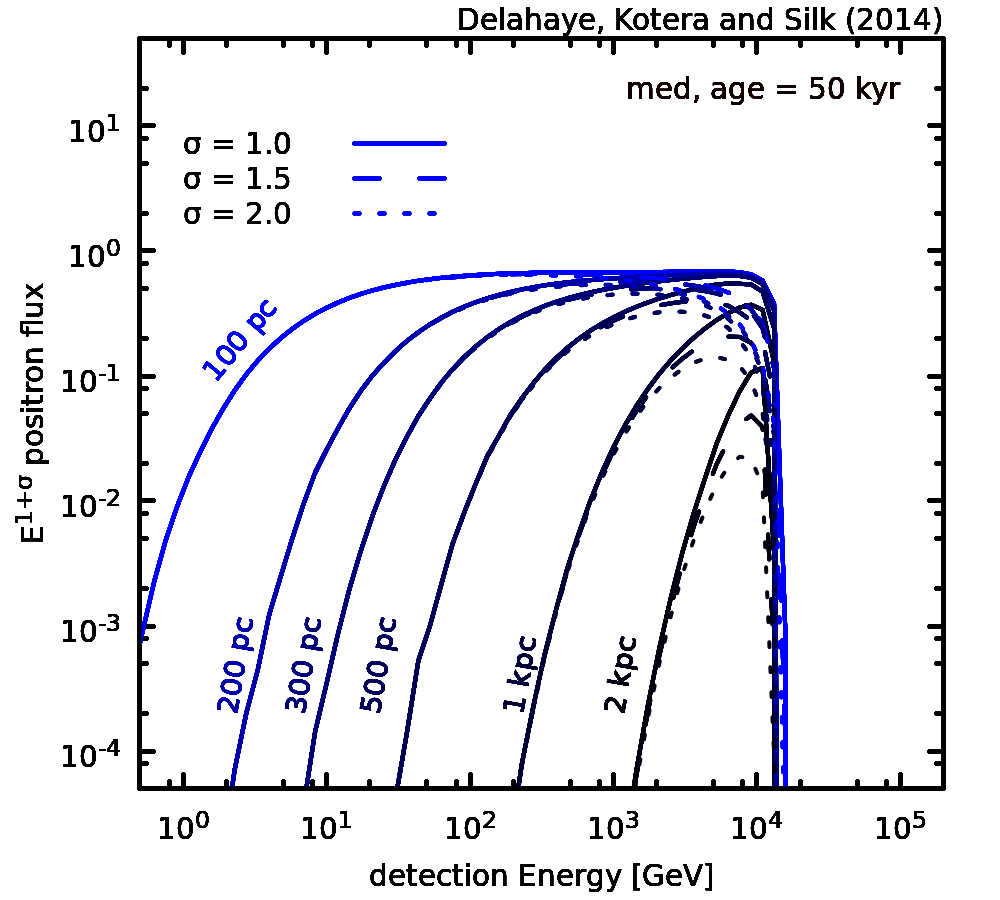}
\caption{Impact of the distance (left panel) and of the age (right panel) of a pulsar on the positron flux received at the Earth. In order to show only the effects of the propagation, the injection energy cut-off has been set to the very high value of 100~TeV for all the cases. Continuous, dashed and dotted lines correspond to an injection respectively of $\sigma=$1, 1.5 and 2. 
The fluxes displayed here are corrected by a factor $E^{1+\sigma}$ to ease the comparison.
It clearly appears that distance has little impact on the shape of the flux at high energies. One should also note that the flux coming from old pulsars drops more sharply, whatever the distance.}
\label{fig-1}
\end{figure*}

If it is quite straightforward that the positron flux coming from Dark Matter cannot reach energies higher than the mass of the Dark Matter particle (or even half this quantity in case of a decaying Dark Matter); if the injection has a rather sharp shape like in the case of annihilation into a electron-positron pair, one can expects the flux after propagation to be quite sharp too. However, the morphology of the positron flux due to a bursting source spatially located, like a pulsar, can be less intuitive. Figure~\ref{fig-1} recalls results from~\cite{Delahaye2010} and compares the flux of positron coming from a single source located 500 pc from us at various times after the injection (left panel) or 500~kyr old located a different distances from us (right panel). We make here the hypothesis that all the cosmic-rays are released at once, for instance at the beginning of the Sedov phase in the case of a pulsar interacting with a supervova remnant. The spectrum of the cosmic-rays at injection in the interstellar medium is a power-law in energy with an exponential cut-off : $\propto E^{-\sigma} \exp\left(E/E_c\right)$.

From Figure~\ref{fig-1}, one can see that older sources can give a sharper drop at high energy and that the distance affects the low energy part of the spectrum but not so much the high energy one. This means that a sharp fall of the positron fraction could be due to a relatively old pulsar within around 2~kpc from the Sun but not necessarily extremely close.

\section*{A sharply falling positron fraction}
\label{sec-2}

\begin{figure*}
\centering
\includegraphics[width=0.9\columnwidth,clip]{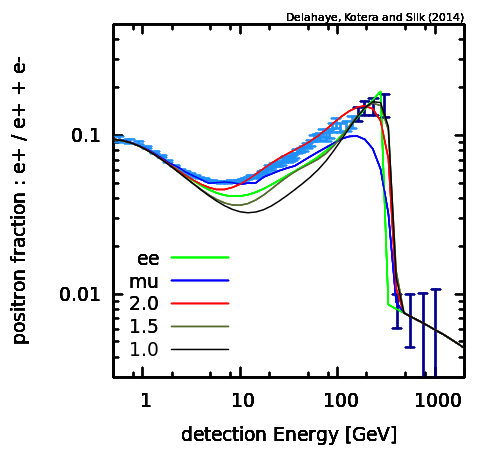}
\includegraphics[width=0.9\columnwidth,clip]{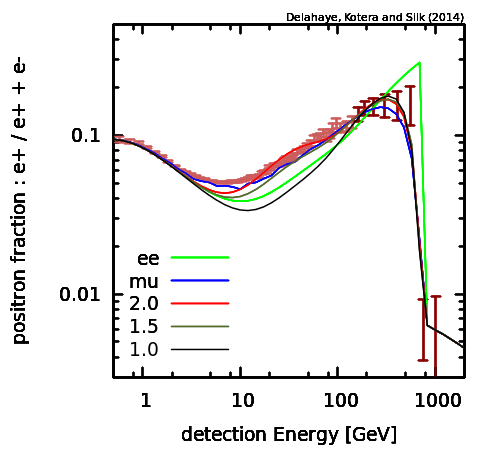}
\caption{Best fit fluxes for the max parameter set. Upper panel for a positron drop at 350~GeV, lower panel at 600~GeV. Data up to 350~GeV is from AMS-02~\citep{Aguilar2013}, above this energy, the bins are mock data. The lines correspond respectively to Dark Matter annihilating into $e^+e^-$ or $\mu^+\mu^-$ or to a pulsar with injection spectrum parameter of 1, 1.5 or 2.
Note that for the pulsar cases, a smooth distribution of far away pulsars, with the same injection spectrum (but a lower cut-off) has been added to reproduce the data at intermediate energies (10 to 150 GeV).}
\label{fig-2}       
\end{figure*}

Let us now consider the hypothesis advertised by the AMS-02 collaboration of a sharply falling positron fraction. We have considered two cases with a sudden drop at 350~GeV and 600~GeV down to the level expected from secondaries (computed as in~\cite{Delahaye2009}). The first case with a drop at 350~GeV is for the discussion's sake only since AMS-02 has now published data up to 500~GeV. Considering that the low energy part could be explained by far away pulsars, we tried to fit the feature of the sharp fall only (the eight last bins that appear darker on Figure~\ref{fig-2}) either with a Dark Matter component or with a single bursting pulsar for which we have considered three possible values of the coefficient $\sigma$: 1.0, 1.5 and 2.0.

As what can see from Figure~\ref{fig-2}, a Dark Matter annihilating into a muon pair does not reproduce well a sharp fall of the positron fraction if the fall happens at too low energy but gives better results if the fraction saturates a little before falling. When Dark Matter annihilates into an electron-positron pair, the fall is too sharp and it is hard to reproduce a fraction that increases slowly and then drops. However, all the pulsars cases can give a good fit. 

Note that in all Dark Matter cases, the annihilation cross-sections (or boost factors) required to fit the data are very high. This is already known for quite some time and raises a large number of issues concerning consistency of such a results with other observations such as anti-protons~\citep{Donato2009}, $\gamma$-rays~\citep{Cirelli:2009dv}, synchrotron emission~\citep{Linden:2011au} etc. 

\begin{figure}
\centering
\includegraphics[width=0.9\columnwidth,clip]{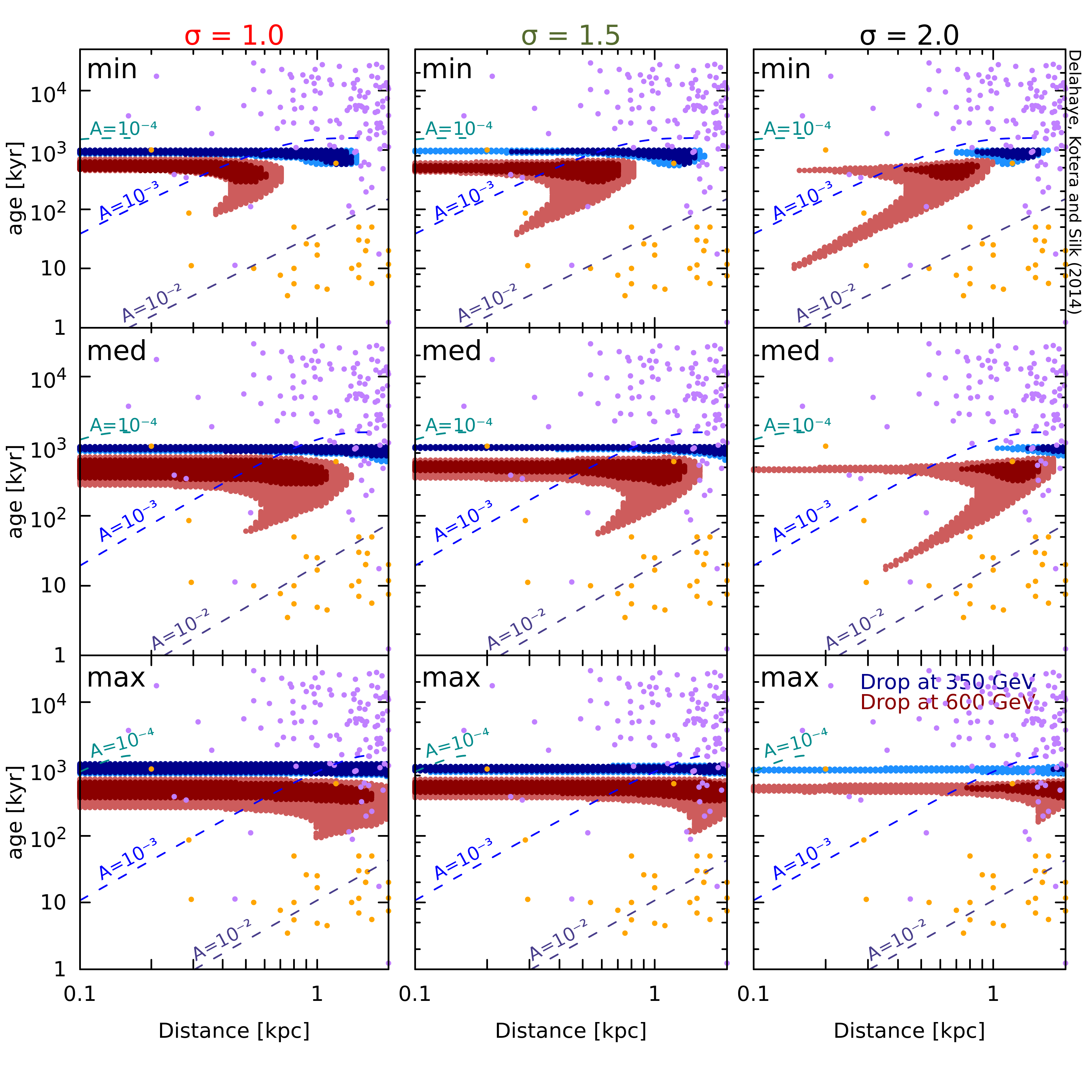}
\caption{Energy cut-off at injection (top) and total energy going to cosmic-ray (bottom) for an injection spectrum $\sigma=1$. The left-hand panels correspond to the min case and a drop at 350~GeV, whereas the right-hand panels are for a drop at 600~GeV and the max propagation parameters.}
\label{fig-3}
\end{figure}

The question is hence, what are the criteria a pulsar would have to fulfil to reproduce a sharply falling positron fraction. Using a fast semi-analytical method for the propagation of cosmic-ray, we have been able to scan a large parameter space. The results of this scan are displayed on Figure~\ref{fig-3}, which shows what ages and distances of the pulsar can accomodate the data, for different injection power-law $\sigma$ (columns) and different propagation parameter sets (lines). The dark area represents paramteres giving a good fit ($\chi^2$/dof $\leq 1$) whereas lighter colors correspond to $2\sigma$ contours. Purple and yellow dots are sources from respectively the ATNF~\cite{Manchester2005} and Green~\cite{Green2009} catalogues.

One can see that for all values of the power-law parameter $\sigma$ it is possible to reproduce a sharply falling positron fraction but that $\sigma=2$ gives a good fit only for a relatively small parameter space, that can be considered as fine tuning. The intuition we had from Figure~\ref{fig-1} is confirmed, the fit procedure prefers a given age but does not care very much about the distance of the source to the Sun. Also, what is important to note is that there are actually a couple of existing sources that are with the correct age and distance parameters.

One cannot conclude anything without also performing an analysis of the energetics involved. For each point of our scan we have computed the amount of energy the source would have to inject into cosmic-rays to give the correct flux at the Earth today. This estimates cannot be extremely precise and depends on the assumptions made for the lowest and highest energies of the cosmic-rays considered, however the values we find amount to around 5 to 10\% of the progenitor supernova ejecta energy. The further the source, the higher the fraction of energy that has to go into cosmic-rays. Our parameter scan favours a relatively old (a few hundred kyr old) close-by source (within $\sim 1$\,kpc), capable of supplying at least ${\cal E}_{\rm tot}\sim 10^{47-48}\,$erg into electrons and positrons, accelerated with a hard spectrum. The discussion concerning the production of such cosmic-rays by a pulsar cannot fit in these short proceedings but the interested reader can refer to \citep{Delahaye2014}.

AMS-02 has also published some limits on the anisotropy of the positron ratio (positron flux divided by negative electron flux). Let us first stress that it is surprising to choose to work with this quantity. Indeed, if there were only a single source of electrons and positrons in the sky, even though most cosmic-rays would come from the same direction, this quantity would be equal to zero since both electrons and positrons would have the same anisotropy. Since electrons are dominating over positrons this quantity is only 20\% smaller than the individual fluxes anisotropies but if the fraction were to increase, this could be problematic. Anyway, we have also computed the anisotropy one would get from a single pulsar responsible for the positron fraction and never found a value that was excluded by the data. An estimate of the positron flux anisotropy $A$ can be read from Figure~\ref{fig-3} (dashed lines).

Most of the parameter space for pulsars compatible with a sharply falling positron fraction is hence compatible with everything we know about cosmic-rays and pulsars.

\section*{Conclusion}

Though the idea that a sharply falling positron fraction would be a proof of Dark Matter is advertised widely by the AMS-02 collaboration, we show here that this is of course not the case. In the contrary, pulsars could explain such a feature in a much more natural way than Dark Matter. However this does not mean that the question is not interesting and one could learn a lot about pulsars if indeed such an unlikely feature were to be observed in the near future.

More precisely, if we really were to observe a sharply falling positron fraction, this would actually teach us that only few pulsars can accelerate electron and positron cosmic-rays. We would then have to understand what are the conditions that make that pulsars can inject cosmic-rays or not. Also, depending of the energy at which the fall is taking place, we could actually determine the age of the pulsar responsible for the feature and look for it in the catalogues and in the sky, allowing to observe more closely the few candidate sources. Finally, by giving us some indication on the injection, this may also help understand what is the precise mechanism that allows these pulsars to inject electron-positron pairs in the interstellar medium.

A sharply falling positron fraction is quite unlikely but if it is observed it would be exciting. Not for the reason that this would prove anything about Dark Matter but because this could potentially teach us a lot about pulsars and cosmic-ray acceleration mechanisms.

\vspace{1cm}
The slides of this presentation are available online~\url{http://www.fysik.su.se/~tdela/SUGAR_2015.html}
\linebreak This work was supported in part by ERC project 267117 (Dark Matters) hosted by Universit\'e Pierre et Marie Curie—Paris 6. K.K. acknowledges financial support from PNHE and ILP.

\bibliography{fraction}

\end{document}